\documentclass[prd,amsmath,amssymb,twocolumn]{revtex4-1}

\usepackage{graphicx}
\usepackage{xcolor}
\usepackage{amsmath}
\usepackage{amssymb}

\usepackage{bm}
\usepackage{bbm}
\usepackage{MnSymbol}

\newcommand{\R}{\mathbb{R}}

\usepackage{xcolor}

\usepackage{color}
 
\begin{document}

\title{An Algebraic Roadmap of Particle Theories  \vspace{2mm}\\ \it Part III:  Intersections\rm}

\author{N. Furey}
\affiliation{$ $\\  Iris Adlershof, Humboldt-Universit\"{a}t zu Berlin,\\ Zum Grossen Windkanal 2, Berlin, 12489 \vspace{2mm} \\ furey@physik.hu-berlin.de \\HU-EP-23/66  }

\begin{abstract}
In this article, we bypass the detailed symmetry breaking pathways established in~\citep{fr1}.  Instead, a direct route from the Spin(10) model to the Standard Model is enabled via a single algebraic constraint.

This single constraint, however, may be reconfigured as a requirement that three $\mathfrak{so}(10)$ actions coincide on a fixed space of multi-vector fermions.  This $\mathfrak{so}(10)\hspace{.5mm}\mapsto \hspace{.5mm}\mathfrak{su}(3)_{\textup{C}}\oplus\mathfrak{su}(2)_{\textup{L}}\oplus\mathfrak{u}(1)_{\textup{Y}}$ breaking (from a three-way intersection) mirrors, in certain ways, the $\mathfrak{so}(8)\hspace{.5mm}\mapsto\hspace{.5mm} \mathfrak{g}_{2}$ breaking (from a three-way intersection) in the context of octonionic triality.  By extending this result to include quaternions and complex numbers, we find that a five-way intersection  breaks $\mathfrak{so}(10)\hspace{.5mm}\mapsto\hspace{.5mm}\mathfrak{su}(3)_{\textup{C}}\oplus\mathfrak{u}(1)_{\textup{Q}}$.  These are the Standard Model's unbroken gauge symmetries, post-Higgs-mechanism.
 \end{abstract}

\maketitle

This article relies on results and notation established in~\citep{fr1}.  It follows  companion papers~\citep{fr1} and \citep{fr2}.

\section{Spin(10) $\mapsto$ Standard Model, directly\label{directly}}

At this time of writing, the Georgi-Glashow model is known to be in a state of tension with experiment, given its unrealized prediction of proton decay, \citep{hisano}, \citep{noscale_su5}.  What if the Pati-Salam and Left-Right Symmetric models are destined to a similar fate?  

In considering such circumstances, we point out the existence of a direct route from the Spin(10) model  to the Standard Model.  

Recall from~\citep{fr1} that $\mathfrak{spin}(10)$ may be realized on a left-handed Weyl spinor as
\begin{equation}\begin{array}{lc}\label{spin10}
\ell_{10}\Psi_{\textup{L}} =& (r_{ab}L_{e_a}L_{e_b} + r'_{mn} L_{\epsilon_m}L_{\epsilon_n} + r_{ma}'' iL_{\epsilon_m}L_{e_a} )\Psi_{\textup{L}},\vspace{2mm}\\
&\mathfrak{spin}(10) \hspace{.7cm}
\end{array}\end{equation}
\noindent for 
$a,b\in\{1,\dots7\}$ with $a\neq b,$ and $m,n\in\{1,2,3\}$ with $m\neq n,$ while $r_{ab},  r'_{mn}, r_{ma}''\in\R.$

Furthermore, standard model symmetries, $\mathfrak{g}_{\textup{SM}} := \mathfrak{su}(3)_{\textup{C}}\oplus\mathfrak{su}(2)_{\textup{L}}\oplus\mathfrak{u}(1)_{\textup{Y}}$, may be realized as
\begin{equation}\begin{array}{rccll}\label{sm}
\ell_{\textup{SM}}\Psi_{\textup{L}} = \big( r_k i \Lambda_k &+& r'_m L_{\epsilon_m}s  &+& irY\big)\Psi_{\textup{L}},\vspace{2mm}\\
\mathfrak{su}(3)_{\textup{C}}&\oplus&\mathfrak{su}(2)_{\textup{L}}&\oplus&\mathfrak{u}(1)_{\textup{Y}}
\end{array}\end{equation}
\noindent for $k\in\{1,\dots 8\}$, $m\in\{1,\dots 3\}$, and $r_k, r'_m, r\in\R$.  The operator $s$ is defined as $s:=\frac{1}{2}(1+iL_{e_7})$.  These $\mathfrak{su}(3)_C$ generators are defined as 
\begin{equation}\begin{array}{lll}\label{su3}
i\Lambda_1:=\frac{1}{2}\left(L_{34}-L_{15} \right) &\hspace{1mm} &
i\Lambda_2:= \frac{1}{2}\left(L_{14}+L_{35} \right)  \vspace{2mm}  \\

i\Lambda_3:=\frac{1}{2}\left(L_{13}-L_{45} \right)&\hspace{1mm} &
i\Lambda_4:= -\frac{1}{2}\left(L_{25}+L_{46} \right)  \vspace{2mm}  \\

i\Lambda_5:=\frac{1}{2}\left(L_{24}-L_{56} \right) &\hspace{1mm} &
i \Lambda_6:=-\frac{1}{2}\left(L_{16}+L_{23} \right)\vspace{2mm}
  \\  
  
i\Lambda_7:= - \frac{1}{2}\left(L_{12}+L_{36} \right)&\hspace{1mm} &  
i \Lambda_8 :=\frac{-1}{2\sqrt{3}}\left(L_{13}+L_{45}  -  2L_{26} \right),  
\end{array}\end{equation}
\noindent where $L_{ij}$ is shorthand for the octonionic $L_{e_i}L_{e_j}$.   Recall that the weak hypercharge generator $iY$ acts as
\begin{equation}iY \Psi_{\textup{L}}:=\left( \frac{1}{6}\left(L_{13} +L_{26}+L_{45} \right) - \frac{1}{2}L_{\epsilon_3}s^*\right)\Psi_{\textup{L}}.
\end{equation}

\begin{figure}[h!]
\begin{center}
\vspace{5mm}
\includegraphics[width=7cm]{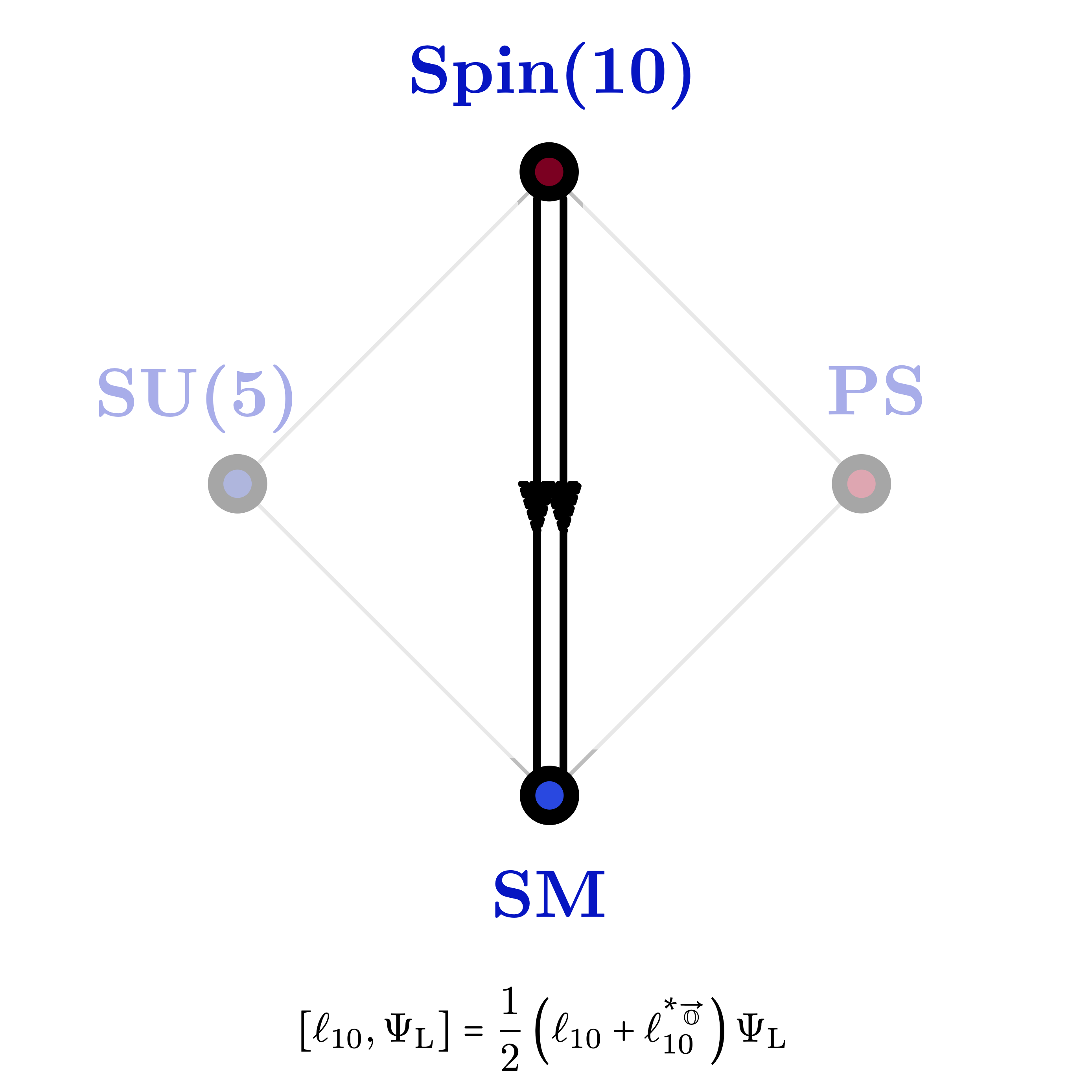}
\caption{\label{Spin10stream}
Spin(10) breaks directly to the Standard Model's symmetries in a single algebraic constraint.}
\end{center}
\end{figure}

Now, it is not difficult to confirm that $\mathfrak{spin}(10)$ of equation~(\ref{spin10}) is sent \it directly \rm to $\mathfrak{su}(3)_{\textup{C}}\oplus\mathfrak{su}(2)_{\textup{L}}\oplus\mathfrak{u}(1)_{\textup{Y}}$ of equation~(\ref{sm}) via a single constraint:
\begin{equation}\label{direct}
\left[ \ell_{10}, \Psi_{\textup{L}}\right] = \frac{1}{2}\left(\ell_{10}+\ell_{10}^{*_{\overrightarrow{\mathbb{O}}}}\right) \Psi_{\textup{L}}.
\end{equation}
\noindent Please see Figure~\ref{Spin10stream}.

We point out in passing that reducing the Lie algebra to its $*_{{\mathbb{D}}}$-invariant subalgebra also affects the spinor space on which it acts.  (Recall that $\mathbb{D}$ represents ${{\overrightarrow{\mathbb{O}}}}$, ${{\overleftarrow{\mathbb{O}}}}$, ${{\overrightarrow{\mathbb{H}}}}$, or ${{\overrightarrow{\mathbb{C}}}}.$)  Namely, $\Psi_{\textup{L}}$ becomes further reducible as 
\begin{equation} \Psi_{\textup{L}} = \Psi_{\textup{L}}^+ \oplus \Psi_{\textup{L}}^-
\end{equation}
\noindent where
\begin{equation}\Psi_L^{\pm} := \frac{1}{2}\left(\Psi_{\textup{L}} \pm \Psi_{\textup{L}}^{*_{{\mathbb{D}}}} \right).
\end{equation}

\section{An overarching theme}
 
It should be apparent for readers that there are two types of algebraic constraints at play in Figure~6 of reference~\citep{fr1}.  Namely, in one direction we find the effects of the multivector condition, and in the other direction, we find the effects of division algebraic reflections.  Might there be an overarching theme to it all?

There is.  In the case of the multivector condition, we found that special unitary symmetries may be isolated from within spin symmetries by \it setting spinor and multivector group actions equal on \rm $\Psi_{\textup{L}}$.  At the Lie algebra level, we have that
\begin{equation} \begin{array}{c}\label{svconstraint} \delta \Psi_{\textup{L}} = [ r_{k_1k_2} \Gamma_{k_1}\Gamma_{k_2}, \hspace{.5mm}\Psi_{\textup{L}}]\hspace{1mm} = r_{k_1k_2} \Gamma_{k_1}\Gamma_{k_2}\Psi_{\textup{L}}, 
\end{array}\end{equation}
\noindent as per equation (18) of~\citep{fr1}.  

Similarly, invariance of the action under division algebraic reflections means that we are \it again setting two different group actions equal on the same fermionic representation space. \rm  Schematically,
\begin{equation} \begin{array}{c}\label{svconstraint} \delta \Psi_{\textup{L}} = \ell \hspace{.5mm}\Psi_{\textup{L}}\hspace{1mm} = \ell^{*_{\mathbb{D}}} \hspace{.5mm}\Psi_{\textup{L}},
\end{array}\end{equation}
\noindent where $\mathbb{D}$ represents ${{\overrightarrow{\mathbb{O}}}}$, ${{\overleftarrow{\mathbb{O}}}}$, ${{\overrightarrow{\mathbb{H}}}}$, or ${{\overrightarrow{\mathbb{C}}}}.$

It is interesting to compare this to the usual spontaneous symmetry breaking scheme employed in the Standard Model.  With the Standard Model, a subgroup is chosen by fixing a (vacuum) element of a particular representation under the original group.  In contrast, here we break symmetries  by allowing the same fermionic space to play different roles, and then insisting on self-consistency.  These two different perspectives need not necessarily be at odds with one another.  The reader is directed to \it Hughes' Higgs \rm in \citep{fh2}, and   to a recent paper  of Hun Jang,~\citep{Hun2023}.

So, an overarching theme that has emerged from viewing the Standard Model in the context of its neighbouring theories is one of \it simultaneous group actions. \rm  

As alluded to earlier in~\citep{fr1}, the idea of simultaneous group actions is not new to group theory.  In fact, it gives us a way to extract the automorphisms of division algebras, e.g. $\mathfrak{aut}(\mathbb{O}),$ from their triality symmetries, e.g. $\mathfrak{tri}(\mathbb{O})$.  

In the remaining sections, we will give an explicit description of this phenomenon in the case of octonionic $\mathfrak{so}(8)$ triality.  That is, we will show how the intersection of three distinct actions of $\mathfrak{so}(8)$ isolates the octonionic derivations, $\mathfrak{der}(\mathbb{O}) =\mathfrak{aut}(\mathbb{O})  = \mathfrak{g}_2$.  

Then, we will draw analogous structures in the context of $\mathfrak{so}(10)$, and will demonstrate that they exhibit a hint of the same theme.  We will show how the intersection of three distinct actions of $\mathfrak{so}(10)$ isolates  $\mathfrak{su}(3)_{\textup{C}}\oplus\mathfrak{su}(2)_{\textup{L}}\oplus\mathfrak{u}(1)_{\textup{Y}}$.  Extending further, we will show how the intersection of five distinct actions of $\mathfrak{so}(10)$ isolates $\mathfrak{su}(3)_{\textup{C}}\oplus\mathfrak{u}(1)_{\textup{Q}}$.  These are the standard model's internal gauge symmetries:  before and after the Higgs mechanism.

\section{Trivializing triality}

It is known that  Spin(8)  happens to have vector, $V$, spinor, $\psi$, and conjugate spinor, $\widetilde{\psi}'$, irreps that are each $8\hspace{.5mm}\R$ dimensional.  Furthermore each of these three 8D representations may be identified with a copy of the octonions.  Here, the tilde on $\psi'$ refers to octonionic conjugation, which sends $e_j\mapsto -e_j$ for $j=\{1,\dots 7\}$, and which reverses the order of multiplication, $\widetilde{xy}=\widetilde{y}\hspace{.8mm}\widetilde{x}$ $\forall x, y, \in \mathbb{O}.$  For further details on the phenomenon of triality, please see~\citep{ms}, \citep{baez}, \citep{mia}.

At the infinitesimal level, \citep{mia}, the vector, spinor, and conjugate spinor actions may be described by
\begin{equation} 
\begin{array}{lcl}
\delta_V f &=& \widehat{d}\hspace{.7mm}f +L_a \hspace{.5mm}f + R_b\hspace{.5mm}f ,\hspace{.5mm} \vspace{2mm}\\
\delta_{\psi}f &=& \widehat{d}\hspace{.7mm}f +L_a \hspace{.5mm}f + R_{a-b}\hspace{.5mm}f ,\vspace{2mm}\\
\delta_{\widetilde\psi}\hspace{.5mm} f &=& \widehat{d}\hspace{.7mm}f +L_{b-a}\hspace{.5mm}f + R_b\hspace{.5mm}f,
\end{array}\end{equation}
\noindent respectively.  Here, $f\in\mathbb{O},$ and $a,b \in \mathfrak{Im}(\mathbb{O})$.  That is, $a=a_je_j$ and  $b=b_je_j$ for $j\in\{1,\dots 7\}$, and $a_j, b_j\in\R.$  The symbol $\widehat{d}$ is shorthand, and represents the octonionic derivations $\mathfrak{aut}(\mathbb{O})=\mathfrak{g}_2$.  Explicitly, 
\begin{equation}\begin{array}{lll}\label{g2}
\widehat{d}&=& r_k \hspace{.5mm}i\Lambda_k \vspace{2mm} \\
&+&  r_9\frac{1}{2\sqrt{3}}\left( L_{15}+L_{34}+2L_{27}\right) + r_{10}\frac{1}{2\sqrt{3}}\left( -L_{14}+L_{35}-2L_{67}\right)\vspace{2mm}\\
&+&  r_{11}\frac{1}{2\sqrt{3}}\left( L_{46}-L_{25}+2L_{17}\right) + r_{12}\frac{1}{2\sqrt{3}}\left( L_{24}+L_{56}-2L_{37}\right)\vspace{2mm}\\
&+&  r_{13}\frac{1}{2\sqrt{3}}\left( -L_{16}+L_{23}+2L_{47}\right) + r_{14}\frac{1}{2\sqrt{3}}\left( L_{12}-L_{36}-2L_{57}\right),
\end{array}\end{equation}
\noindent where $k\in\{1,\dots 8\}$, and $r_i \in \R$ for $i\in\{1,\dots 14\}.$

\begin{figure}[h!]
\begin{center}
\includegraphics[width=8cm]{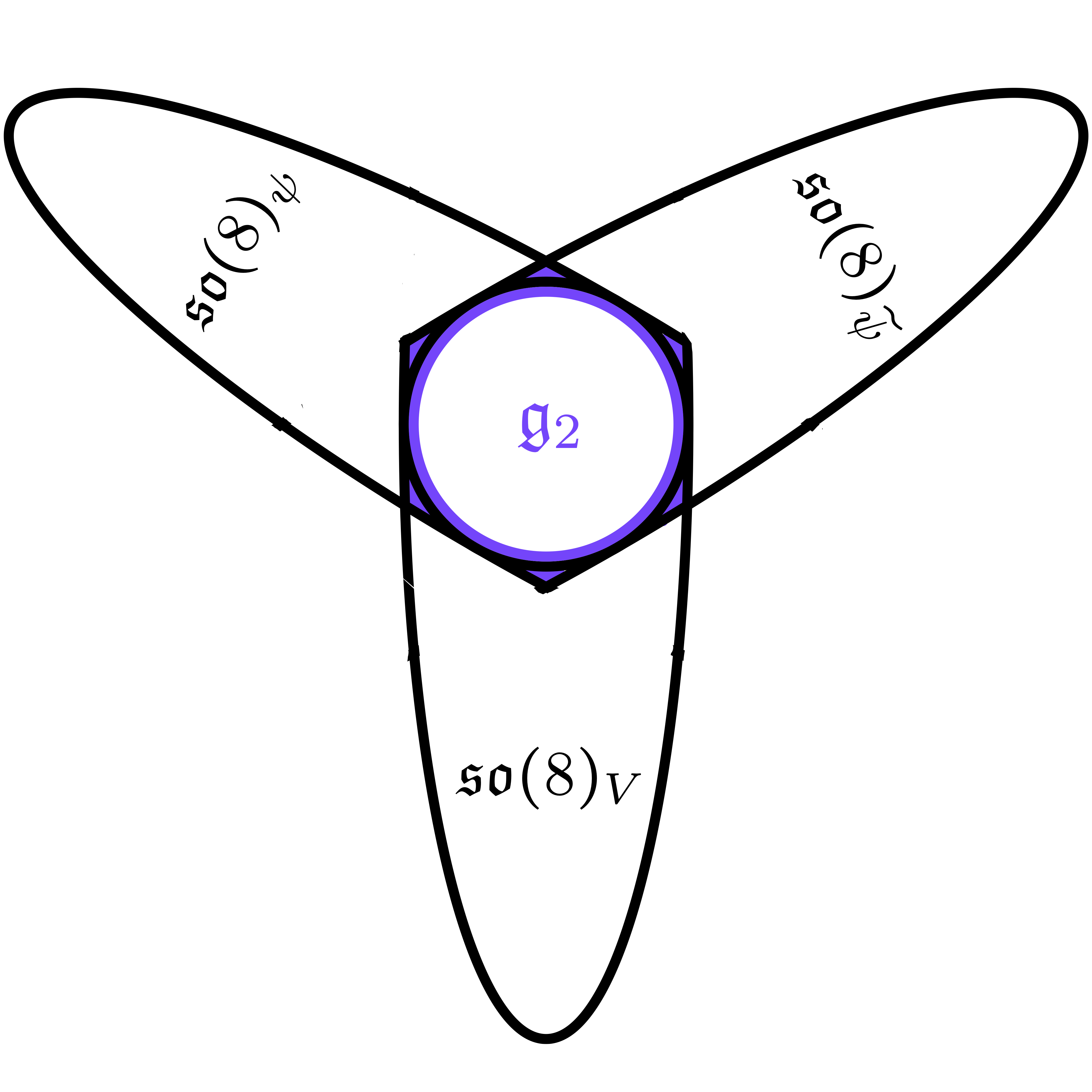}
\caption{\label{tri}
The spinor action, conjugate spinor action, and vector action of $\mathfrak{so}(8)$ coincide on the octonionic derivations, $\mathfrak{aut}(\mathbb{O})=\mathfrak{g}_2$.}
\end{center}
\end{figure}

Now, given that the vector, spinor, and conjugate spinor irreps can each be identified with a copy of the octonions, one might wonder what happens when we ask that their three Lie algebra actions coincide.  It is straightforward to confirm that these three $\mathfrak{so}(8)$ actions coincide on $\widehat{d}$, the octonionic derivations $\mathfrak{der}(\mathbb{O})=\mathfrak{aut}(\mathbb{O})=\mathfrak{g}_2.$  Please see Figure~\ref{tri}.

\section{A parallel in particle physics}

Curiously enough, there happens to be a similar pattern in particle physics that bears some resemblance to the three-way intersection of Figure~\ref{tri}.  It results from an alternative perspective on the constraint equation~(\ref{direct}) that we obtained in Section~\ref{directly}:
\begin{equation}\label{direct2}
\left[ \ell_{10}, \Psi_{\textup{L}}\right] = \frac{1}{2}\left(\ell_{10}+\ell_{10}^{*_{\overrightarrow{\mathbb{O}}}}\right) \Psi_{\textup{L}}.
\end{equation}

\begin{figure}[h!]
\begin{center}
\includegraphics[width=8cm]{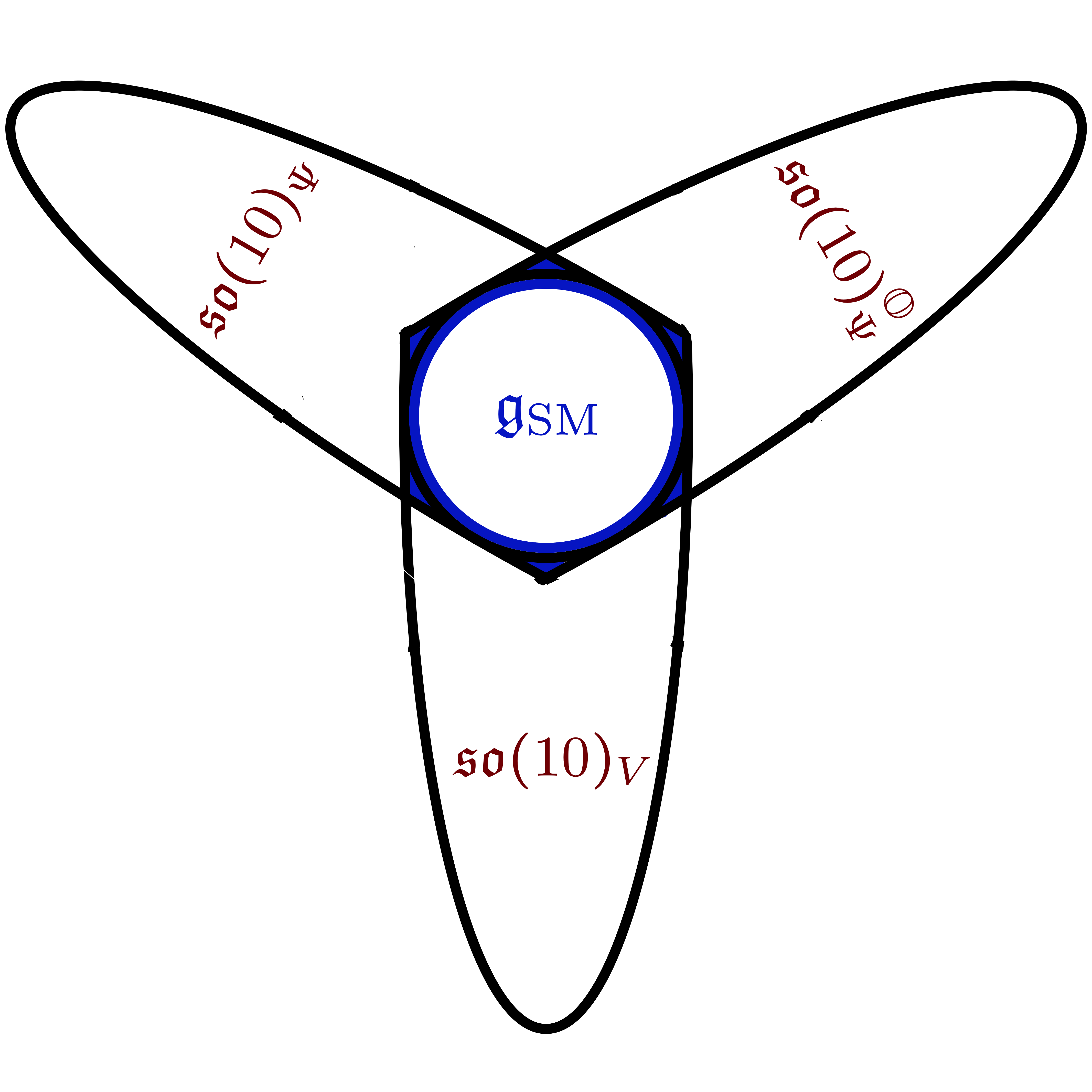}
\caption{\label{trism}
The spinor action, octonionic-conjugate spinor action, and multi-vector action of $\mathfrak{so}(10)$ coincide on the Standard Model's internal symmetries, $\mathfrak{g}_{\textup{SM}}=\mathfrak{su}(3)_{\textup{C}}\oplus \mathfrak{su}(2)_{\textup{L}}\oplus \mathfrak{u}(1)_{\textup{Y}}$.}
\end{center}
\end{figure}

Much like the three-way constraint that sends $\mathfrak{so}(8) \mapsto \mathfrak{g}_{2}$, we find a three-way constraint that sends $\mathfrak{so}(10) \mapsto \mathfrak{g}_{\textup{SM}}$.  For all $\Psi_{\textup{L}}$,
\begin{equation} \begin{array}{c}\label{sm3way} 
\delta   \Psi_{\textup{L}}=\vspace{2mm} \\
\left[ \ell_{10}, \hspace{.5mm} \Psi_{\textup{L}}\right] =  \ell_{10}\Psi_{\textup{L}} = \ell_{10}^{*_{{\overrightarrow{\mathbb{O}}}}}\Psi_{\textup{L}}
  \vspace{2mm}\\
\Rightarrow \vspace{2mm} \\
\ell_{10}\hspace{2mm} \mapsto \hspace{2mm}\ell_{\textup{SM}}\vspace{2mm} \\
\mathfrak{so}(10) \hspace{2mm}\mapsto  \hspace{2mm}\mathfrak{su}(3)_{\textup{C}}\oplus \mathfrak{su}(2)_{\textup{L}}\oplus \mathfrak{u}(1)_{\textup{Y}}.
\end{array}\end{equation}
\noindent Please see Figure~\ref{trism}.  In the figure, $\mathfrak{so}(10)_V$ refers to the action $\left[ \ell_{10}, \hspace{.5mm} \Psi_{\textup{L}}\right],$ while $\mathfrak{so}(10)_{\Psi}$ refers to the action $\ell_{10} \Psi_{\textup{L}},$  and $\mathfrak{so}(10)_{\Psi}^{\mathbb{O}}$ refers to the action $\ell_{10}^{*_{{\overrightarrow{\mathbb{O}}}}}\Psi_{\textup{L}}.$ 

Those familiar with Figure~6 of reference~\citep{fr1} might now wonder:  What occurs when such intersections are made to include not only octonionic reflections, but quaternionic and complex reflections as well.  

\begin{figure}[h!]
\begin{center}
\includegraphics[width=8cm]{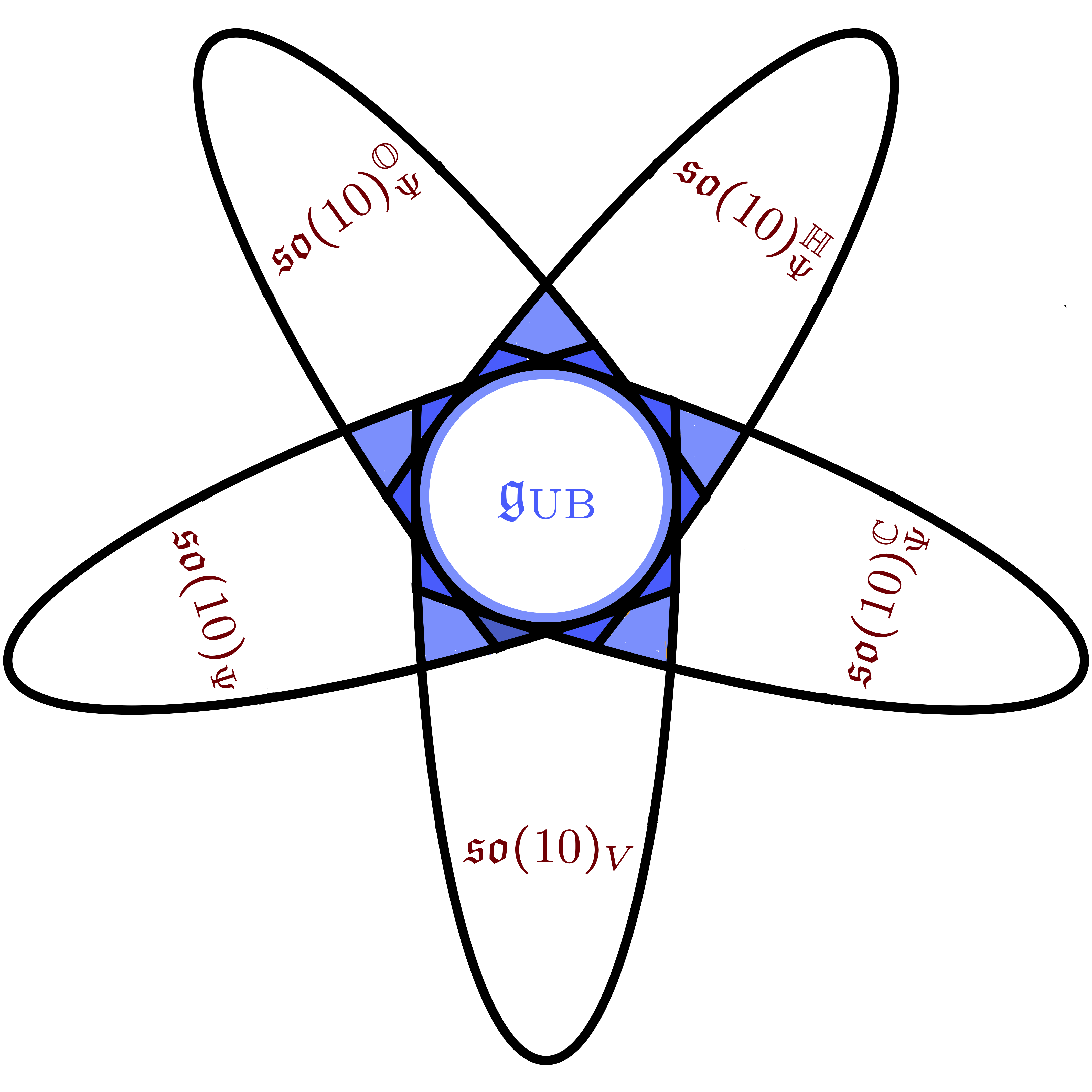}
\caption{\label{5UB}
The spinor action, the octonionic-conjugate spinor action, quaternionic-conjugate spinor action, complex-conjugate spinor action, and multivector action of $\mathfrak{so}(10)$ coincide on the Standard Model's unbroken symmetries, $\mathfrak{g}_{\textup{UB}}=\mathfrak{su}(3)_{\textup{C}}\oplus\mathfrak{u}(1)_{\textup{Q}}$.}
\end{center}
\end{figure}

In this case, we find a five-way intersection that maps $\mathfrak{so}(10) \mapsto \mathfrak{g}_{\textup{UB}} := \mathfrak{su}(3)_{\textup{C}}\oplus\mathfrak{u}(1)_{\textup{Q}}$, the Standard Model's internal gauge symmetries that remain after the Higgs mechanism. For all $\Psi_{\textup{L}},$
\begin{equation} \begin{array}{c}\label{sm3way} 
\delta   \Psi_{\textup{L}}=\vspace{2.5mm} \\
\left[ \ell_{10}, \hspace{.5mm} \Psi_{\textup{L}}\right] =  \ell_{10}\Psi_{\textup{L}} = \ell_{10}^{*_{{\overrightarrow{\mathbb{O}}}}}\Psi_{\textup{L}}= \ell_{10}^{*_{{\overrightarrow{\mathbb{H}}}}}\Psi_{\textup{L}} = \ell_{10}^{*_{{\overrightarrow{\mathbb{C}}}}}\Psi_{\textup{L}}
  \vspace{2mm}\\
\Rightarrow \vspace{2mm} \\
\ell_{10}\hspace{2mm} \mapsto \hspace{2mm}\ell_{\textup{UB}}\vspace{2mm}\\
\mathfrak{so}(10)\hspace{2mm}\mapsto\hspace{2mm}\mathfrak{su}(3)_C\oplus\mathfrak{u}(1)_Q.
\end{array}\end{equation}
\noindent Here, $\ell_{\textup{UB}}$ is defined as 
\begin{equation}\begin{array}{lrcl}
\ell_{\textup{UB}}\Psi_{\textup{L}} = &\big( r_j i \Lambda_j &+& r'iQ\big)\Psi_{\textup{L}},
\end{array}\end{equation}
\noindent where $r'\in\R,$ and $iQ$ turns out to be the electric charge generator,
\begin{equation}iQ:= \frac{1}{6}\left( L_{13}+L_{26}+L_{45}\right)-\frac{1}{2}L_{\epsilon_3}.
\end{equation}
\noindent Please see Figure~\ref{5UB}.  In the figure,  $\mathfrak{so}(10)_{\Psi}^{\mathbb{H}}$ furthermore refers to the action $\ell_{10}^{*_{{\overrightarrow{\mathbb{H}}}}}\Psi_{\textup{L}}$,  and $\mathfrak{so}(10)_{\Psi}^{\mathbb{C}}$  refers to the action $\ell_{10}^{*_{{\overrightarrow{\mathbb{C}}}}}\Psi_{\textup{L}}$.

 \section{Conclusion}
 
In this article, we identified a single algebraic constraint that sends $\mathfrak{so}(10)$ to the Standard Model's internal gauge symmetries, $\mathfrak{g}_{\textup{SM}} = \mathfrak{su}(3)_{\textup{C}}\oplus\mathfrak{su}(2)_{\textup{L}}\oplus\mathfrak{u}(1)_{\textup{Y}}$.

Reworking this constraint, however, allowed us to view it as a coincidence between three inequivalent $\mathfrak{so}(10)$ actions.  Such a coincidence constraint is already familiar from octonionic triality, where three $\mathfrak{so}(8)$ actions coincide on $\mathfrak{aut}(\mathbb{O})=\mathfrak{g}_2$.

We extended the idea further to include not only octonionic, but also quaternionic and complex conjugate type reflections.  In this case, the five-way intersection of $\mathfrak{so}(10)$ actions results in $\mathfrak{g}_{\textup{UB}} := \mathfrak{su}(3)_C\oplus\mathfrak{u}(1)_Q$ - those Standard Model  gauge symmetries surviving the Higgs mechanism.

\begin{acknowledgments}  
The author is  grateful for discussions, feedback, and encouragement from John Baez, Suk\d{r}ti Bansal, John Barrett, Latham Boyle,  Hilary Carteret, Olaf Hohm, Mia Hughes, Kaushlendra Kumar, Agostino Patella, Beth Romano, Matthias Staudacher, Shadi Tahvildar-Zadeh, Carlos Tamarit, Andreas Trautner, and Jorge Zanelli.

This work was graciously supported by the VW Stiftung Freigeist Fellowship, and Humboldt-Universit\"{a}t zu Berlin.



\end{acknowledgments}

\medskip

\end{document}